\documentclass[twocolumn,superscriptaddress,nofootinbib, floatfix, amsfonts, prd]{revtex4-1}
\usepackage{amsmath}
\usepackage{bm}
\usepackage{graphicx}
\usepackage{subfigure}
\usepackage{float}
\usepackage{mathrsfs}
\usepackage{footmisc}
\usepackage{multirow}
\usepackage{graphicx}
\usepackage{booktabs, ctable}
\usepackage{soul,xcolor}
\usepackage{xcolor, ulem, color, framed}
\usepackage{amssymb}
\usepackage{setspace}
\usepackage[colorlinks, linkcolor=red,anchorcolor=green,citecolor=blue]{hyperref}

\graphicspath{{./figs-Nov/}}

\begin{document}
\begin{spacing}{1.0}

\title{Machine learning approach to analyze heavy quark diffusion coefficient in relativistic heavy-ion collisions}

\author{Rui Guo}
\affiliation{Data science, Washington University in St. Louis, MO 63105, USA}

\author{Yonghui Li}\email{yonghui.li@tju.edu.cn}
\affiliation{Department of Physics, Tianjin University, Tianjin 300354, China}

\author{Baoyi Chen}\email{baoyi.chen@tju.edu.cn}
\affiliation{Department of Physics, Tianjin University, Tianjin 300354, China}

\date{\today}

\begin{abstract}
The diffusion coefficient of heavy quarks in the deconfined medium is examined in this research using a deep convolutional neural network (CNN) trained with data from relativistic heavy ion collisions involving heavy flavor hadrons. The CNN is trained using observables such as the nuclear modification factor $R_{AA}$ and the elliptic flow $v_2$ of non-prompt $J/\psi$ from B-hadron decay in different centralities, where B meson evolutions are calculated using the Langevin equation and the Instantaneous Coalescence Model. The CNN outputs the parameters characterizing the temperature and momentum dependence of the heavy quark diffusion coefficient. By inputting the experimental data of non-prompt $J/\psi$ $(R_{AA}, v_2)$ from various collision centralities into multiple channels of the well-trained network, we derive the values of the diffusion coefficient parameters. Additionally, We evaluate the uncertainty in determining the diffusion coefficient by taking into account the uncertainties present in the experimental data $(R_{AA}, v_2)$,  which serve as inputs to the deep neural network.

\end{abstract}
\maketitle

\section{Introduction}

In recent years, there has been rapid development in deep learning methods, which are increasingly being widely applied in industry and scientific research. In particular, deep learning methods are used to handle high-dimensional data and uncover patterns, such as in image recognition and more. In the realm of scientific research, deep learning has already found applications in many aspects of physics~\cite{Radovic:2018dip,Carleo:2019ptp,Mehta:2018dln}. In theoretical research in high-energy nuclear physics, an increasing number of studies are utilizing deep learning methods to analyze computational data from theoretical models, such as the study of the equation of state of QGP matter~\cite{Pang:2016vdc}, dynamical evolutions of QGP~\cite{Huang:2018fzn}, identifying spinodal clumping in high energy nuclear collisions~\cite{Steinheimer:2019iso}.

In the Relativistic Heavy-Ion Collider (RHIC) and the Large Hadron Collider (LHC)~\cite{Bazavov:2011nk,Shuryak:1980tp,Aoki:2006we}, a new kind of deconfined state, called Quark-Gluon Plasma (QGP), is predicted and produced.
The properties such as initial energy density and the coupling strength of QGP are typically studied through the final-state light hadron spectra~\cite{NA49:2002pzu,Song:2007ux,Shen:2011eg,Song:2010mg} or the distribution of heavy flavor particles~\cite{Yan:2006ve,Liu:2010ej,Rapp:2008tf,Qin:2015srf,Andronic:2003zv,Matsui:1986dk}. For heavy flavor particles, their distribution is initially influenced by the cold nuclear effect, and then they will interact with the QGP medium, resulting in energy loss~\cite{Altenkort:2023oms,Qin:2007rn,Zhang:2003wk,Guo:2000nz}.
Numerous models have been established to account for the various effects mentioned above~\cite{Cao:2016gvr,He:2014cla,Ke:2018jem,Chen:2021akx,Chen:2017duy,Akamatsu:2018xim}, aiming to ultimately predict the nuclear modification factor and anisotropic flows of open heavy flavor hadrons. 
These models investigate the energy loss mechanisms of heavy quarks in the QGP from different perspectives. Relevant nuclear modification factors and collective flows of D mesons or B mesons have also been experimentally measured by STAR~\cite{STAR:2017kkh}, ALICE~\cite{ALICE:2013olq,ALICE:2014qvj,ALICE:2015ccw} and CMS~\cite{CMS:2017uuv,CMS:2022gvy} Collaborations, which are closely related to the energy loss process of heavy quarks. 
As it is not straightforward to explain the experimental observables of $R_{AA}$ and $v_2(p_T)$ at the same time with a simple value of the diffusion coefficient $D_s 2\pi T$, a more realistic expression with temperature and momentum dependence is needed. Due to the complex
process involving the heavy quark diffusion and hydrodynamic evolution, it is necessary to employ deep
neural networks to analyze the relationship between the diffusion coefficient and experimental observables. CNN has been proven to be suitable
for analyzing high-dimensional datasets and quantify the value of the diffusion coefficient considering multiple hot and cold nuclear matter effects.

In previous studies, Bayesian statistical analysis has been employed to analyze the experimental data of soft particles~\cite{Bernhard:2016tnd,Auvinen:2017fjw,Novak:2013bqa,Pratt:2015zsa}, quantitatively extract the diffusion coefficient with experimental observables of charmed hadrons in heavy-ion collisions~\cite{Xu:2017obm}. Although there is abundant experimental data on D mesons, our main focus is on studying the evolution of B mesons. This is because the diffusion coefficient is directly related to the drag term in the Langevin equation, which arises from elastic scattering processes rather than the medium-induced gluon radiation. Therefore, as the mass of the heavy quark increases, the contribution from elastic scattering processes becomes more significant in the energy loss process, while the contribution from gluon radiation is relatively weaker. 
In this work, we treat the $R_{AA}$ and $v_2$ of non-prompt $J/\psi$ from B-meson decay as inputs of the CNN and the parameters in the diffusion coefficient as outputs of the network. After training the neural network with supervision, the inputs of the neural network are selected from values within the error bars of the experimental data points in order to generate the corresponding diffusion coefficient with some uncertainty.

This paper is organized as follows: in Section II, we introduce the Langevin plus Instantanuous Coalescence model (LICM) to generate datasets of heavy flavor evolutions with different values of the parameters, which will be used as trainning datasets of the CNN model. In Section III, the values of the shadowing factor, temperature and momentum dependence of the diffusion coefficient are quantitatively extracted based on the experimental data of non-prompt $J/\psi$ from B-meson decay. A final summary is given in Section IV.

\section{Frameworks}
\subsection{Generating datasets with LICM}

Bottom quarks are produced in the hard scatterings of nucleon partons. We adopted the the fixed-order plus
next-to-leading log formula (FONLL)~\cite{FONLL,FONLL-2} and NNPDF30NLO PDF set~\cite{NNPDF:2014otw} to calculated the initial momentum distribution of bottom quarks in nucleon-nucleon collisions. Due to the fact that Pb-Pb collisions can be regarded as a superposition of nucleon-nucleon collisions, accompanied by cold nuclear matter effects, the initial momentum distribution of bottom quarks in Pb-Pb collisions can be considered as the momentum distribution in pp collisions multiplied by a shadowing factor.
The nuclear shadowing factor is calculated with the EPS09 NLO package~\cite{Eskola:2009uj} in 5.02 TeV Pb-Pb collisions. 
The production of bottom quarks primarily arises from binary collision processes, hence the spatial distribution of bottom quarks in nuclear collisions is proportional to the distribution of binary collisions, $dN_{b\bar b}/d{\bf x}_T\propto T_A({\bf x}_T-{\bf b}/2) T_B({\bf x}_T+{\bf b}/2)$~\cite{Yang:2023rgb}. 
Here, $T_{A(B)}=\int dz \rho_{A(B)}({\bf x}_T, z)$ represents the thickness functions of the two nuclei. The nucleon distribution $\rho({\bf x}_T, z)$ in the nucleus follows a Woods-Saxon distribution.

After the production of bottom quarks, they propagate within the high-temperature QGP medium accompanied by energy loss. The energy loss of bottom quarks in the QGP is primarily attributed to scattering processes between bottom quarks and thermal partons, as well as medium-induced parton radiation. Considering that the mass of the bottom quark is much larger than the typical temperature of the medium, the momentum change during each interaction of bottom quarks in the medium is relatively small, which can be regarded as Brownian motion. Consequently, the momentum evolution of bottom quarks can be described using the Langevin equation,
\begin{align}
\label{eq-lan}
    {d{\bf p}\over dt} = -\eta_D {\bf p}+{\bf \xi}+{\bf f}_g. 
\end{align}
On the right-hand side, the first two terms represent the drag and noise terms, which come from the elastic collisions with thermal light partons. The drag coefficient is defined as $\eta_D(p)=\kappa/(2TE_b)$, where the energy of the bottom quark is given by $E_b=\sqrt{m_b^2 +{p}^2}$.  The mass of the bottom quark is taken to be 
$m_b=4.75$ GeV. $\kappa$ represents the momentum-diffusion coefficient. It is related to the spatial diffusion coefficient $D_s$ via $D_s \kappa = 2T^2$. In the noise term $\bf \xi$, the time correlation and momentum dependence are both neglected for simplicity, where $\bf \xi$ is treated as a Gaussian shaped white noise satisfying the relation, 
\begin{align}
\langle \xi^{i}(t)\xi^{j}(t^\prime)\rangle =\kappa \delta ^{ij}\delta(t-t^\prime).
\end{align}
The third term ${\bf f}_g=-d{\bf p}_g/dt$ with ${\bf p}_g$ being the momentum of the emitted gluon, represents the recoil force on the bottom quark from the emitted gluon. The number of emitted gluons in a small time interval $t\sim t+\Delta t$ is~\cite{Cao:2013ita}, 
\begin{align}
\label{gluon-spec}
P_{\rm rad}(t,\Delta t) = \langle N_g(t, \Delta t)\rangle = \Delta t \int dx d k_T^2
{dN_g\over dx dk_T^2dt}.
\end{align}
Here $x=E_g/E_b$ represents the ratio of energy carried by gluons radiated from bottom quarks.
$dN_g/dxdk_T^2dt$ is the 
the spectrum of emitted gluon from higher twist calculation~\cite{Zhang:2003wk,Guo:2000nz}. $k_T$ is the transverse momentum 
of the gluon. The position of heavy quark is updated in each time step as ${\bf x}(t+\Delta t)= {\bf x}(t) + {\bf p}/E_b\cdot \Delta t$. 

Heavy quarks are randomly generated based on the initial spatial and momentum distributions, and propagate in the QGP with energy loss described with Eq.(\ref{eq-lan}). When bottom quarks diffuse and move into certain regions where QGP local temperature is low, heavy quarks undergo hadronization by combining with light quarks to form B mesons or by combining with an anti-heavy quark to form quarkonium. In the high-momentum region, the production of B mesons from bottom quarks is predominantly through fragmentation processes, while in the intermediate and low-momentum regions, it is mainly through the coalescence process. In this study, we primarily focus on bottom quarks with transverse momentum $p_T\le 15$ GeV/c. Therefore, we utilize a coalescence model to describe the hadronization process of bottom quarks into B mesons,
{\small
\begin{align}
\label{eq-Dcoal}
{dN_M\over d{\bf p}_M}
=\int {d{\bf p}_1\over (2\pi)^3} {d{\bf p}_2\over (2\pi)^3}
{dN_1\over d{\bf p}_1} {dN_2\over d{\bf p}_2 }
f_M^W({\bf q}_r)\delta^{(3)}({\bf p}_M -{\bf p}_1-{\bf p}_2), 
\end{align}  }
The momentum distribution of B meson $dN_M/d{\bf p}_M$ is proportional to the distributions of bottom quarks $dN_1/d{\bf p}_1$ and also the thermal light quarks $dN_2/d{\bf p}_2$. Heavy quark distribution is given by the Langevin equation, while thermal light quark is taken as a Fermi distribution. Their coalescence probability is determined by the Wigner function $f_M^W({\bf q}_r)$, which can be obtained via the Weyl transform of B meson wave function. In principle, the complete Wigner function $f^W({\bf q}_r, {\bf x}_r)$ provides constraints on the relative distance and relative momentum between two particles for the formation of a bound state. The spatial constraint becomes crucial and can significantly reduce the coalescence probability when the two particles are rare in the QGP, as observed in the case of charmonium coalescence process. However, in the case of a B meson composed of one heavy and one light anti-quark, with a plentiful number of light quarks in the QGP, we assume that the heavy quark can readily find a light quark in proximity, thus satisfying the spatial constraint. By integrating over the spatial part of the complete Wigner function, we simplify it to a normalized Gaussian function $A_0 \exp(-q_r^2 \sigma^2)$, where $A_0$ represents the normalization factor. The width of the Gaussian function is connected with the root-mean-square radius of B meson, $\sigma^2= {4\over 3}{(m_1+m_2)^2\over m_1^2 +m_2^2}
\langle r^2\rangle_B $~\cite{Greco:2003vf}, with the value to be 
$\sqrt{\langle r^2\rangle_B}=0.43$ fm~\cite{Yang:2023rgb}. $m_1$ is the bottom quark mass, while the thermal mass of the light quark is approximated to be $m_2=0.3$ GeV, used in the coalescence process. 
${\bf q}_r=(E_2^{\rm cm}{\bf p}_1^{\rm cm}- E_1^{\rm cm}{\bf p}_2^{\rm cm})/(E_1^{\rm cm}+E_2^{\rm cm})$ is the relative momentum between bottom quark and the light quark in the center of mass (COM) frame. ${\bf p}_1^{\rm cm}$ and ${\bf p}_2^{\rm cm}$ are the momenta of the bottom quark and the light quark 
in the COM frame. The delta function ensures the momentum conservation in the coalescence process ${\bf p}_M={\bf p}_1+{\bf p}_2$. As the phase transition between QGP and hadronic gas crossover at LHC energies, we perform the coalescence process at the critical temperature $T_c=150$ MeV. The time and spatial evolutions of bulk medium has been well described with the hydrodynamic equations. We employ the MUSIC package to give the information of hot medium at 5.02 TeV Pb-Pb collisions~\cite{Schenke:2010rr,Schenke:2010nt, Chen:2021akx}. The local temperatures of the medium vary with coordinates and time, and these variations will be incorporated into the Langevin equation.
After the hadronization, B mesons continue diffusion in the hadronic gas with a different value of the diffusion coefficient, and decay into non-prompt $J/\psi$ after the kinetic freeze-out at the temperature $T_{\rm fo}=120$ MeV.

Recently, Lattice QCD calculations present the new calculations of the spatial diffusion coefficient of heavy quarks at different temperatures~\cite{Altenkort:2023oms}, which is found to be smaller than the previous quenched lattice QCD~\cite{Altenkort:2020fgs,Brambilla:2020siz} and recent phenomenological estimates~\cite{Liu:2016ysz,ALICE:2021rxa,Xu:2017obm,Scardina:2017ipo}. This conclusion is also observed in other theoretical results~\cite{Andreev:2017bvr,Casalderrey-Solana:2006fio,Altenkort:2023eav}. This prompts us to reexamine the relationship between experimental measurements of heavy quarks and the diffusion coefficient. In the high temperature and momentum regimes, the diffusion coefficient can be calculated through perturbative QCD~\cite{Combridge:1978kx,Caron-Huot:2008dyw,Caron-Huot:2007rwy}. However, this calculation is not sufficient to simultaneously explain the $R_{AA}$ and $v_2$ of open heavy flavor particles measured in nuclear collisions~\cite{Das:2015ana}, suggesting that non-perturbative processes play an indispensable role in the temperature and momentum dependence of the diffusion coefficient. In Bayesian statistical analysis, a parameterized form for the diffusion coefficient is proposed, including a linear temperature dependence term and a perturbative QCD term, such as $D_s2\pi T\propto A(p)(\alpha +\beta T)+(1-A(p))8\pi/({\hat q}/T^3)$~\cite{Xu:2017obm}. The first term represents the contribution from non-perturbative processes, while the second term represents the contribution from perturbative processes. 
$\hat q$ is the heavy quark transport coefficient, which is calculated by elastic scatterings between heavy and light quarks~\cite{Baier:1996sk}. The spatial diffusion coefficient from lattice QCD calculations is in the $p=0$ case. In this work, we introduce a concise formula to consider the temperature and momentum dependence in the spatial diffusion coefficient, 
\begin{align}
\label{eq-diffc}
    D_s2\pi T= [\alpha +\beta({T\over T_c}-1)]\times({m_Q\over E_Q})^\gamma. 
\end{align}
The temperature and momentum dependence are encoded in the terms $T/T_c$ and $m_Q/E_Q$. $m_Q$ and $E_Q=\sqrt{m_Q^2 +p_Q^2}$ are the mass and energy of heavy quark. The parameter $\alpha$ represents the value of $D_s2\pi T$ at the critical temperature with the momentum $p=0$. The parameter $\beta$ and $\gamma$ controls the degree of temperature and momentum dependence. In the hadronic gas, the coupling strength between B- and D- mesons and the medium becomes much smaller. Their contribution on the $R_{AA}$ and $v_2$ of open heavy flavor particles are limited. In hadronic gas, the mean value of the spatial diffusion coefficient of B-meson is approximated to be $D_s^{M}2\pi T=9$ before the kinetic freeze-out of B-meson in $0.8T_c<T<T_c$~\cite{He:2012df}.

\subsection{Deep neural networks}
\begin{figure}[htb]
\centering
\includegraphics[width=0.48\textwidth]{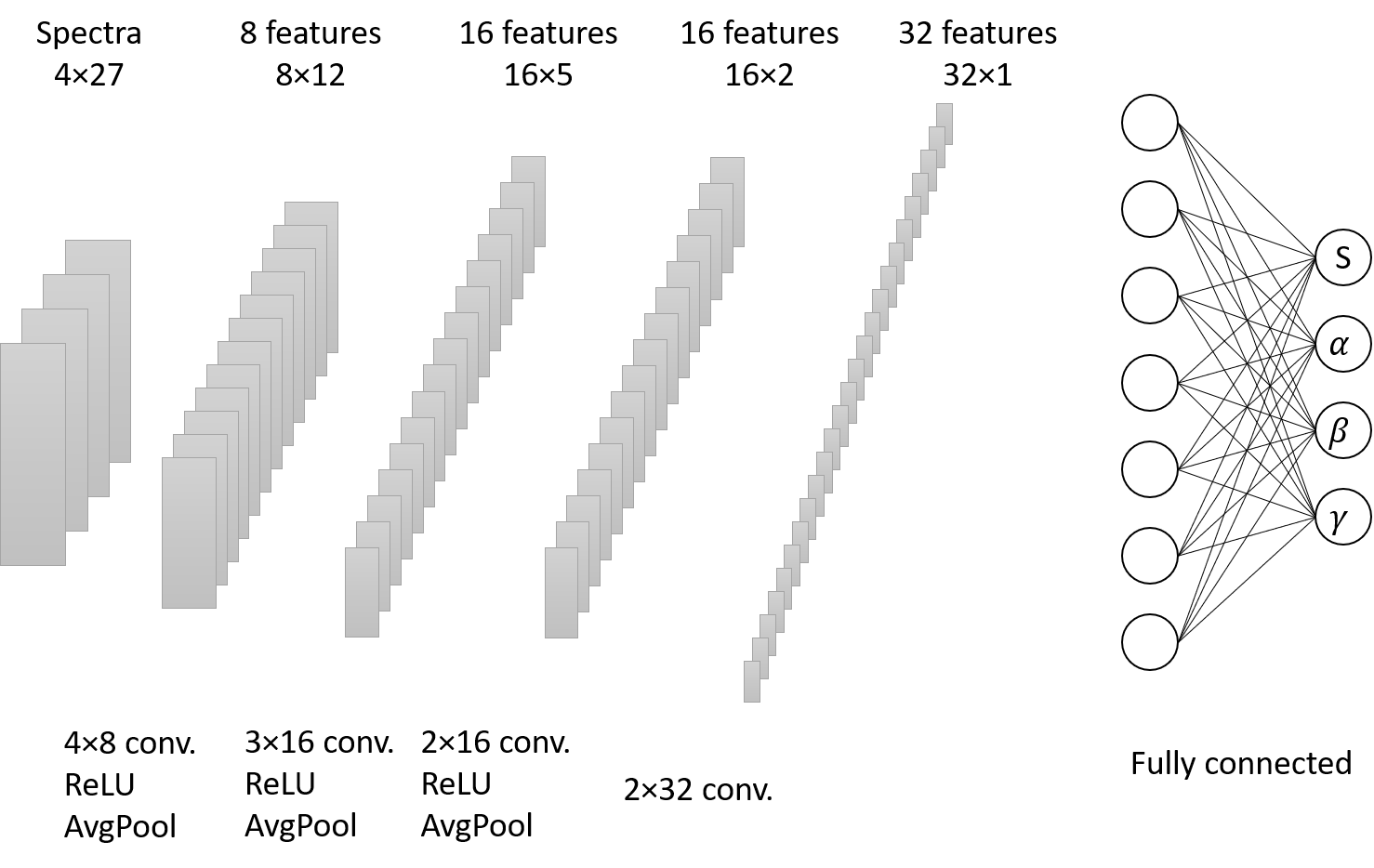}
\caption{ Schematic figure to show the structure of the convolutional neural network. $R_{AA}(p_T)$ in three centralities and one $v_2(p_T)$ are taken as four channels of the input layer, while parameters related to the spatial diffusion coefficient are the output. 
}
\label{lab-cnn}
\end{figure}

In the dynamical evolution of bottom quarks, our theoretical model produces a wide range of 
$R_{AA}$ and $v_2$ values for non-prompt 
$J/\psi$ in 5.02 TeV Pb-Pb collisions by varying the shadowing factor denoted as 
$S$, and the three parameters 
$(\alpha, \beta, \gamma)$ in Eq.(\ref{eq-diffc}).
This dataset will serve as the training data for the CNN. Experimental measurements have been conducted to determine the nuclear modification factor 
$R_{AA}(p_T)$ for non-prompt 
$J/\psi$ in three different centralities, as well as the elliptic flow coefficient 
$v_2(p_T)$. In our approach, we treat the three centralities of 
$R_{AA}(p_T)$ as separate channels, while 
$v_2(p_T)$ acts as an additional channel in the input layer of the CNN. The output layer of the CNN incorporates the corresponding parameter values in the diffusion coefficient and the shadowing factor, which are considered as labels for the input data. Theoretical calculations based on the LICM model establish a mapping relationship between the parameter combination values 
$(S, \alpha, \beta, \gamma)$ and the experimental observables 
$(R_{AA}, v_2)$. 
Figure \ref{lab-cnn} provides a graphical representation of the network structure, which consists of four hidden layers including Average pooling layers and fully connected layers. The ReLU activation function is chosen for these layers. The three output channels related to the diffusion coefficient parameters are projected to have positive values, while the channel associated with the shadowing effect is projected within the range of 0-1 using the Sigmoid function.

To generate the training dataset, we randomly select parameter values within the regions specified in Table~\ref{tab:train} using the Langevin model. Due to the significant uncertainty surrounding the shadowing factor $S$ in model calculations, which can notably impact the final observables of the B meson, we consider the shadowing factor as a parameter to be optimized within the deep neural network. The values of the shadowing factor are constrained within the range of 0.6 to 1.0. For 5.02 TeV Pb-Pb collisions, the values of $S$ used to generate the training dataset are randomly selected from the range of 0.6 to 1.0. The spatial diffusion coefficient 
$D_s 2\pi T$ at the critical temperature 
$T_c$ is selected within the range of 
$2.0\le \alpha \le 6.0$, while the values of 
$\beta$ and $\gamma$ characterizing the temperature and momentum dependence are chosen within the ranges of 
$0\le \beta\le 8.0$ and 
$0\le \gamma \le 1.0$, respectively. We generate 20K events as the training and validation datasets, where each event corresponds to one combination of parameters.
The performance of the neural network is influenced by both the size of the training dataset and the structure of the deep neural network. This type of uncertainty has been examined in previous studies~\cite{Barnard:2016qma} and will be partially investigated in this work by varying the size of the training datasets. However, such uncertainties are not as significant as those arising from the error-bars of the experimental data points of heavy flavor particles used as input for the CNN.

\begin{table}[htbp]
  \centering
  \caption{The samples of the parameters used in the training dataset}
  \label{tab:train}
  \begin{tabular}{|c|c|}
    \hline
     parameters  & region \\
     \hline
     shadow factor & $S\in [0.6, 1.0]$ \\
     \hline
     $D_s 2\pi T$ & $\alpha\in [2.0,6.0]$ \\
     with &  \\
     T-dependence & $\beta\in[0.0,8.0]$ \\
     $p_T$-dependence & $\gamma\in[0.0, 1.0]$ \\
     \hline
       \end{tabular}
\end{table}

\section{results and analyse}
\begin{figure}[htb]
\includegraphics[width=0.4\textwidth]{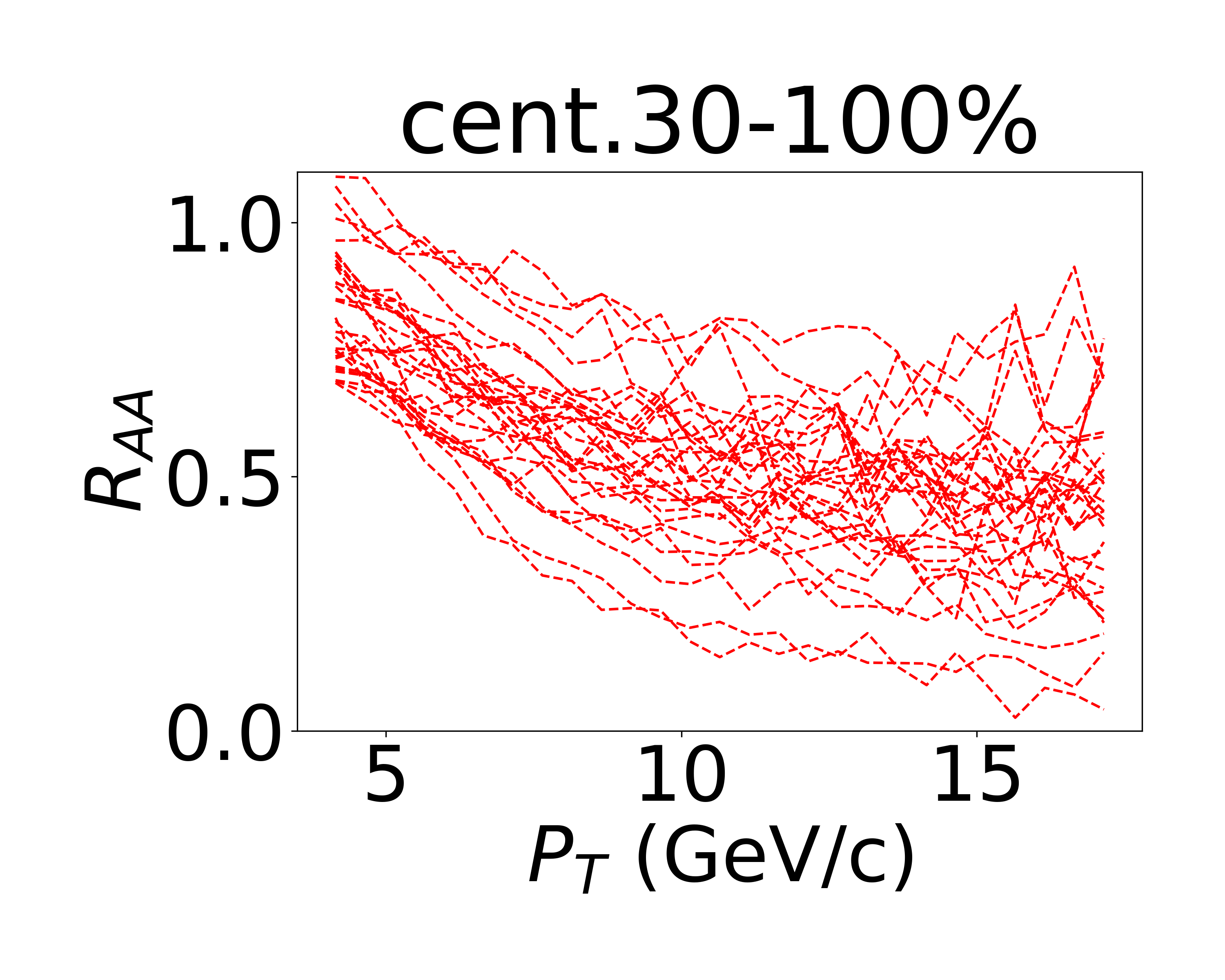}
\caption{ Some events randomly selected from one channel of the training dataset. 
The lines represent the nuclear modification factors of non-prompt $J/\psi$ calculated with different values of the parameters in the centrality 30-100\% in 5.02 TeV Pb-Pb collisions. 
}
\label{lab-raa-example}
\end{figure}

In the previous sections, we introduce the theoretical model to generate the training dataset for the CNN by varying the values of the parameters in the model. 
We plot some events $(R_{AA}(p_T), v_2(p_T))$ randomly selected from one channel of the training datasets. They are shown in Fig.\ref{lab-raa-example}. The lines can cover the experimental data, which indicates that the training range of the model encompasses the distribution of experimental data. 
The model can be effectively applied to analyze the experimental data. As experimental data points about $R_{AA}$ are located in $p_T\ge 2$ GeV/c and $v_2$ is located in $p_T\ge 4$ GeV/c, We truncate the training data by only retaining data with 
$p_T$ values above 4 GeV/c. This ensures that the shape of the training data aligns as closely as possible with the experimental data, making it easier to incorporate into the input, please see Fig.\ref{lab-raa-example}.

\begin{figure}[!htb]
\includegraphics[width=0.45\textwidth]{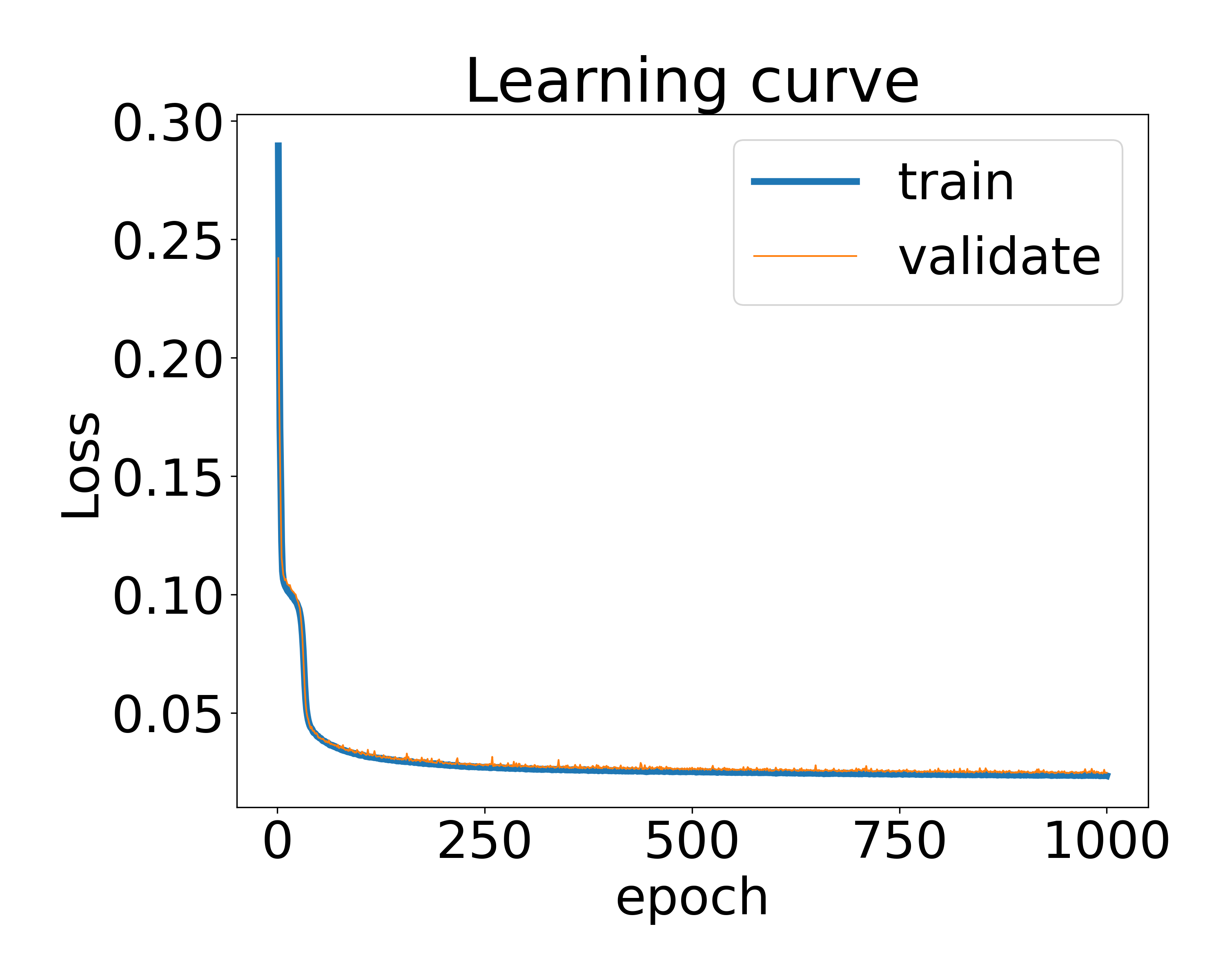}
\caption{The loss of the CNN model as a function of the training epochs. The loss of the model calculated with the training datasets and the validation datasets are plotted respectively.
}
\label{lab-loss}
\end{figure}

We partition 70\% of the total datasets as the training data, while the remaining 30\% of the datasets serve as the validation data. By treating 
$(R_{AA},v_2)$ as inputs for the CNN and the parameter values as labels, we can calculate the loss of the CNN for both the training and validation datasets. The learning curve shown in Fig. \ref{lab-loss} demonstrates that the loss decreases to below 5\% after 250 training epochs.
Notably, the loss of the CNN for the training datasets closely aligns with the loss observed when using the validation datasets. This indicates that the CNN model does not exhibit significant overfitting or underfitting.

\begin{figure}[!htb]
\includegraphics[width=0.22\textwidth]{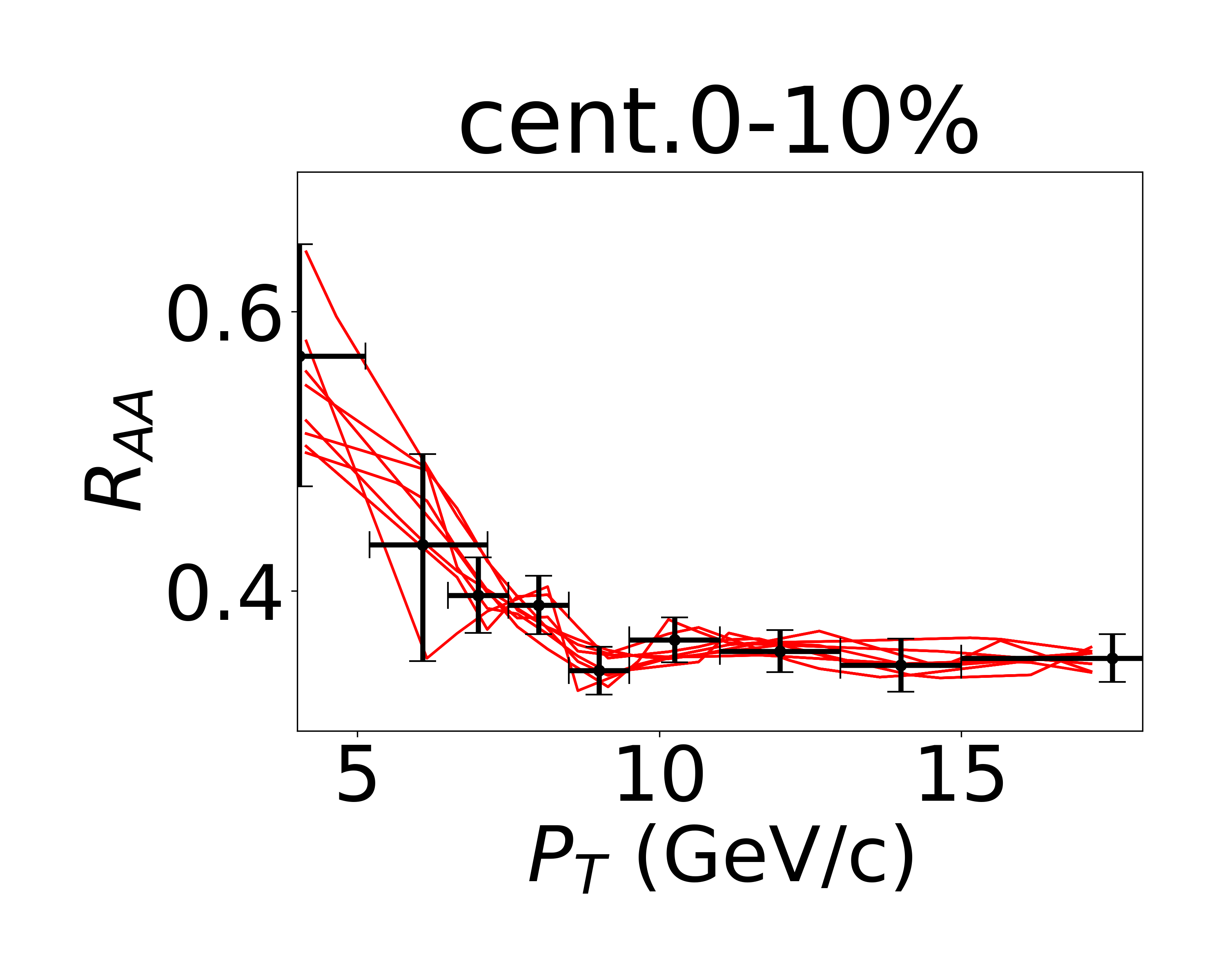}
\includegraphics[width=0.22\textwidth]{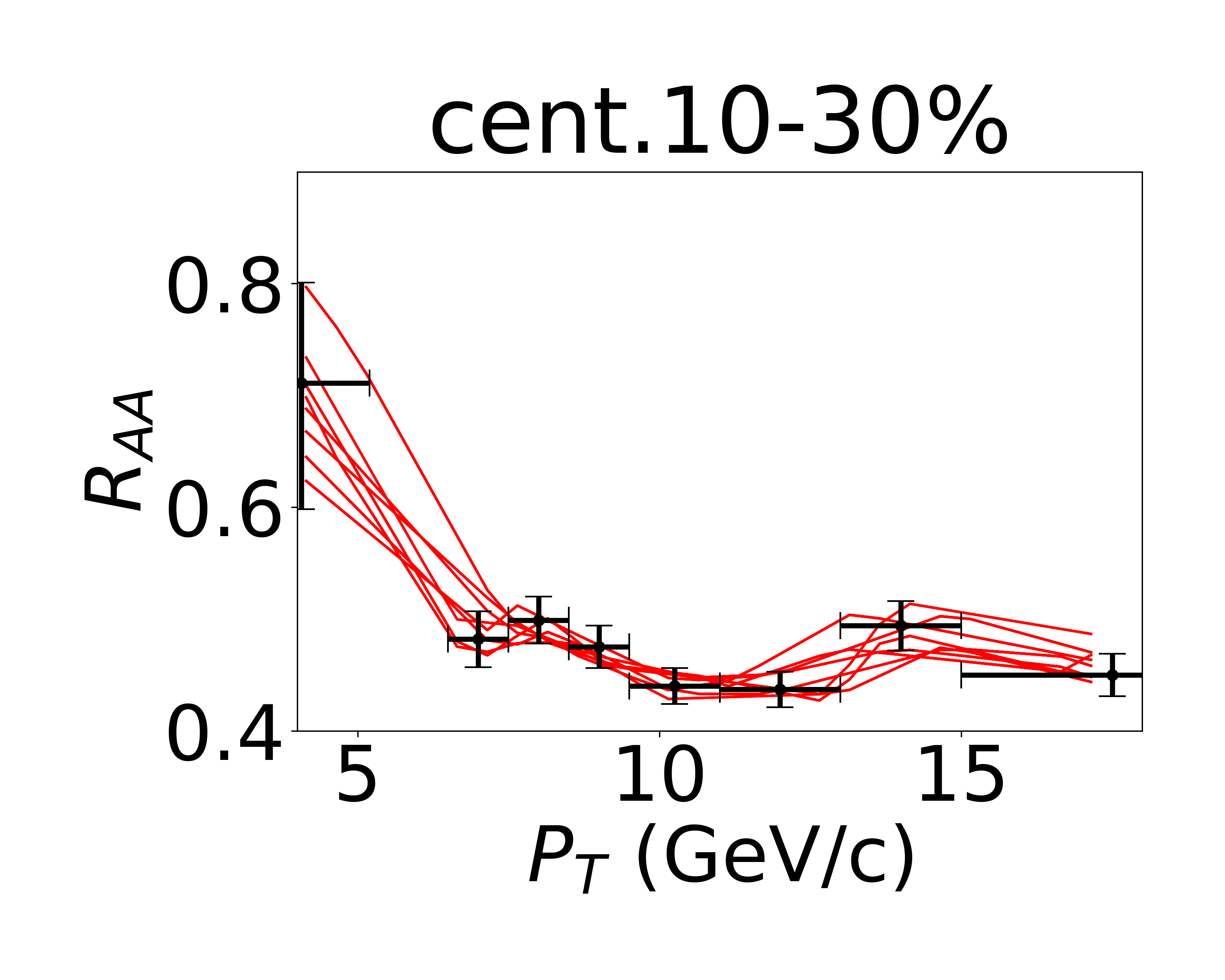}
\includegraphics[width=0.22\textwidth]{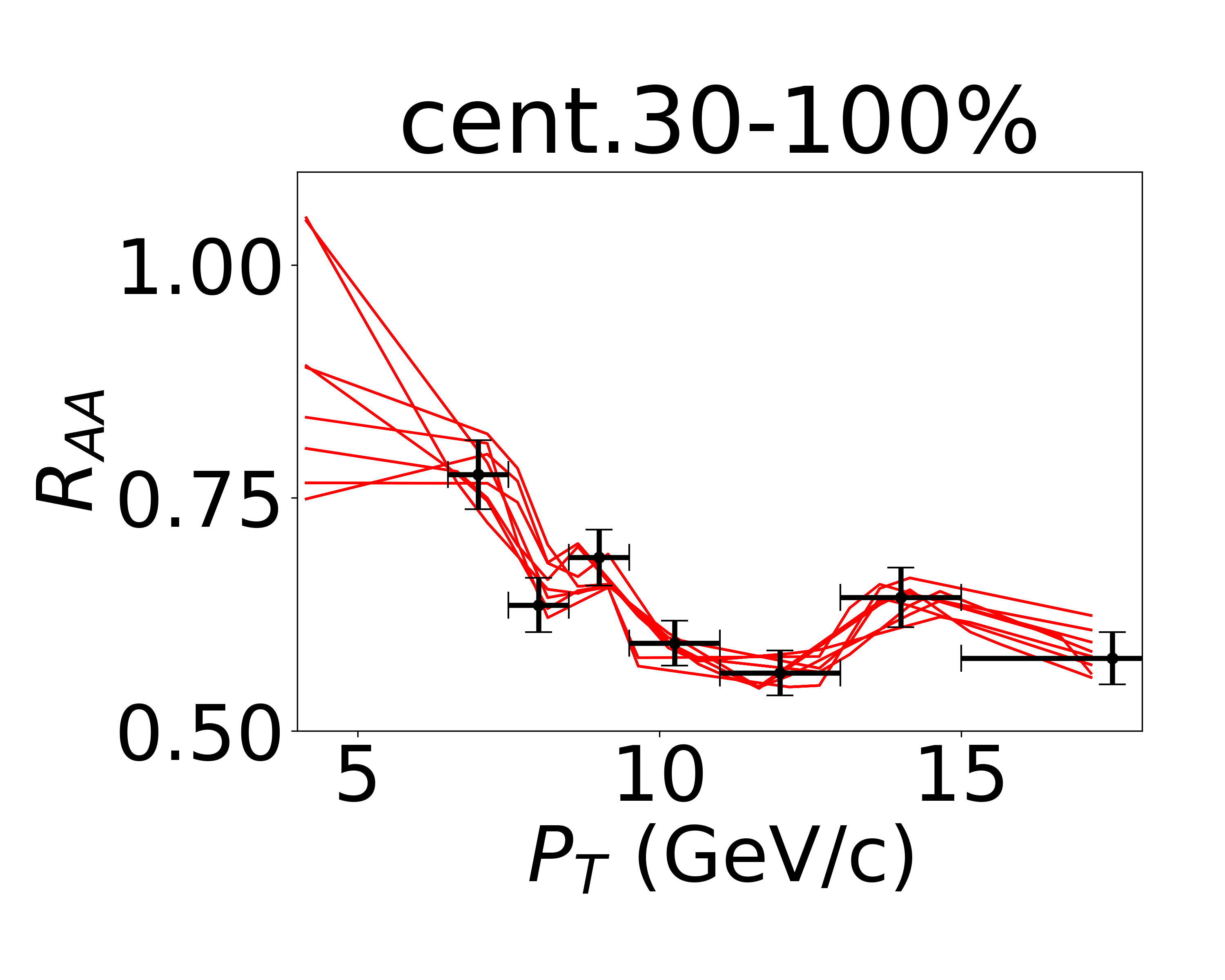}
\includegraphics[width=0.22\textwidth]{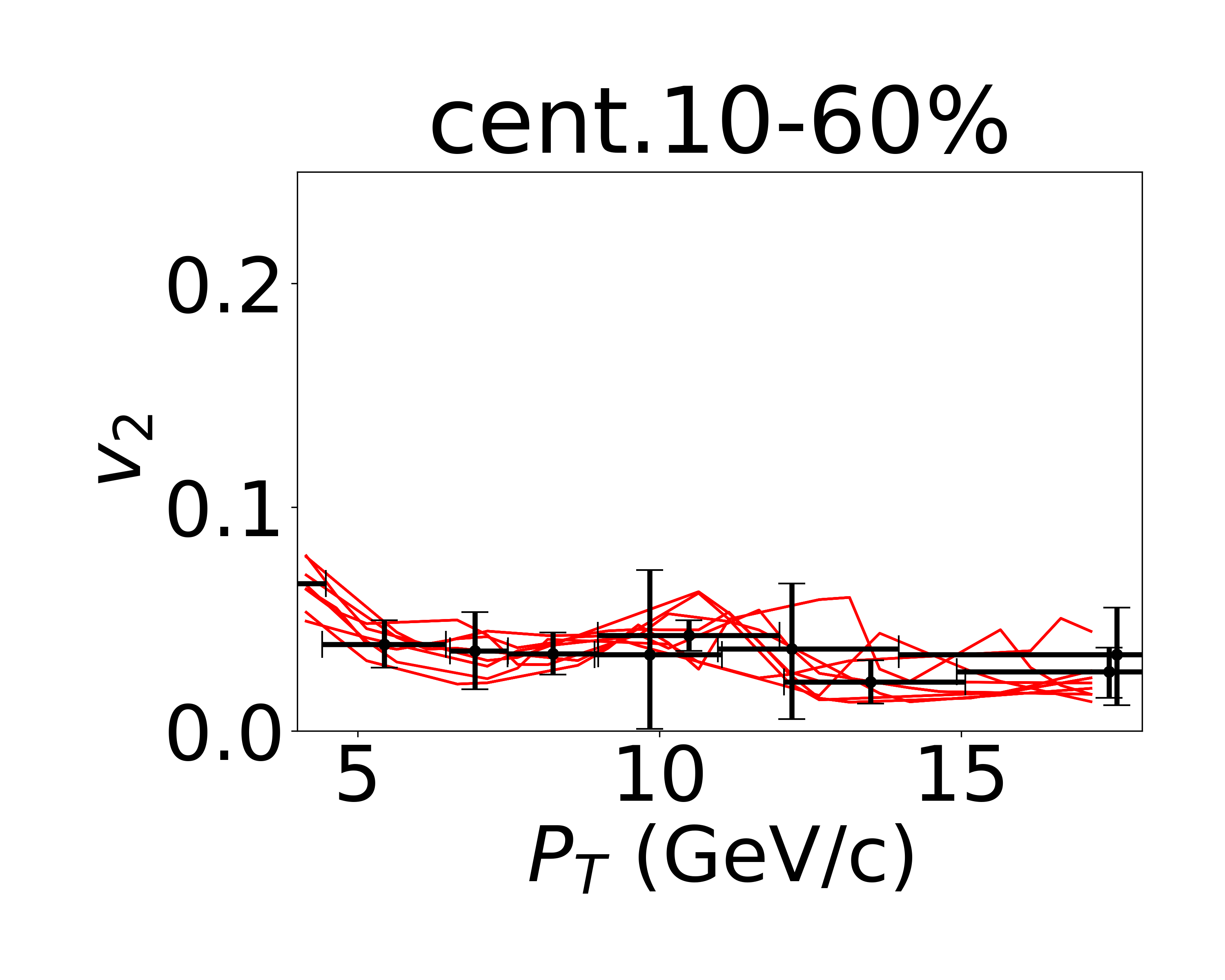}
\caption{ To consider the error-bars of the experimental data about non-prompt $J/\psi$ in 5.02 TeV Pb-Pb collisions, we sample the values of $R_{AA}$ and $v_2$ within the error-bars of each data points, and take them as inputs of the deep neural network. Some lines representing random events are plotted in the figures. Four figures represents four channels of the network. The experimental data are cited from CMS Collaboration~\cite{CMS:2017uuv,CMS:2022gvy}.
}
\label{lab-exp-input}
\end{figure}

Using the well-trained CNN model, we feed the experimental data points into the neural network. Considering the presence of error bars in the experimental data points (as depicted in Fig. \ref{lab-exp-input}), we sample within the error-bars associated with the experimental data points. These samples are then utilized as inputs to the neural network, allowing us to account for the impact of experimental uncertainties on the diffusion coefficient parameters.
To capture the influence of experimental uncertainty, we randomly generate 10K samples within the range of experimental error-bars around the data points, as illustrated in Fig. \ref{lab-exp-input}.

\begin{figure}[!htb]
\includegraphics[width=0.4\textwidth]{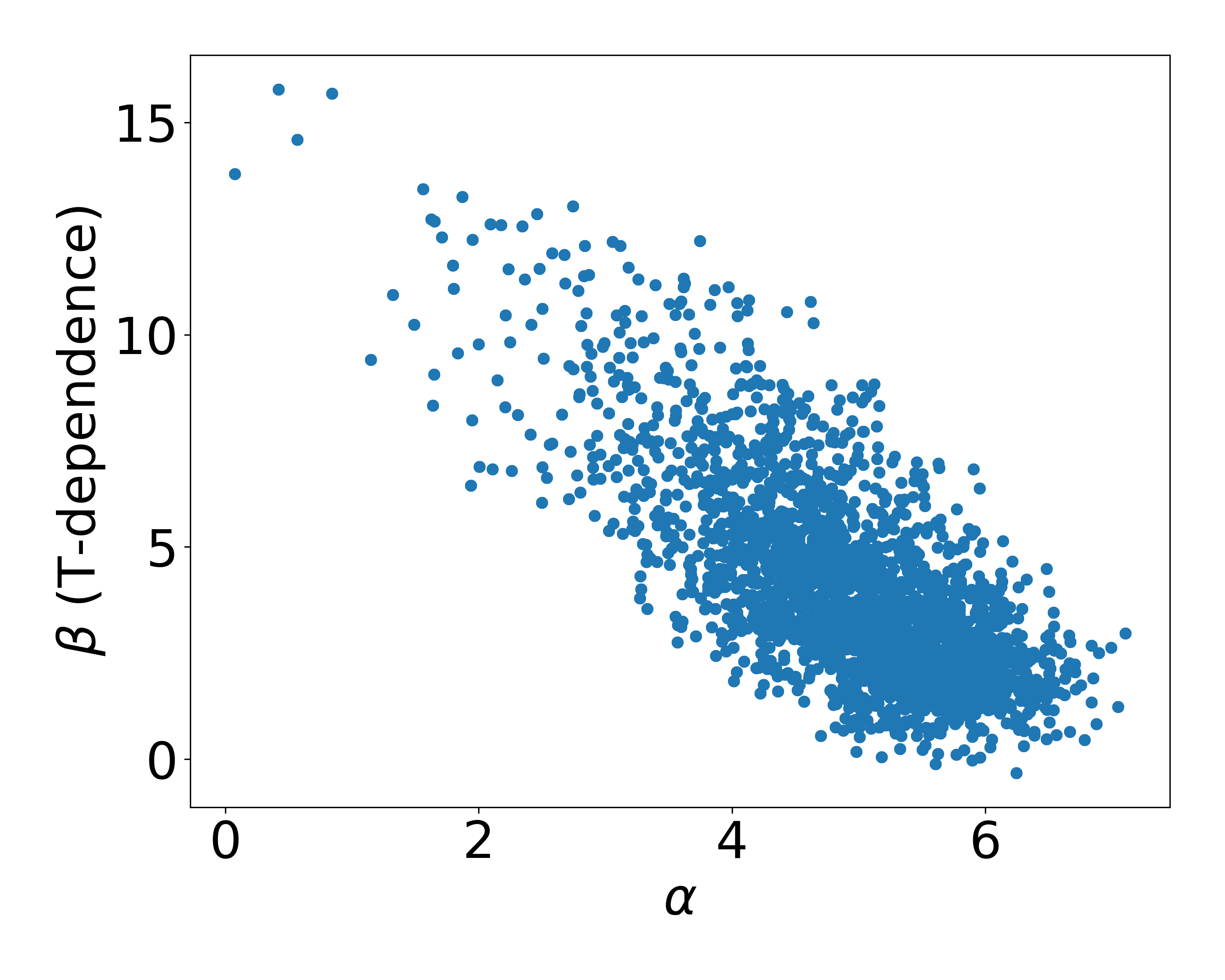}
\caption{ The distribution of parameter values $(\alpha, \beta)$ from the CNN. The x-axis represents $\alpha$, while the y-axis represents $\beta$. Each point represent one event. 
}
\label{lab-param-alpha}
\end{figure}

\begin{figure}[!htb]
\includegraphics[width=0.39\textwidth]{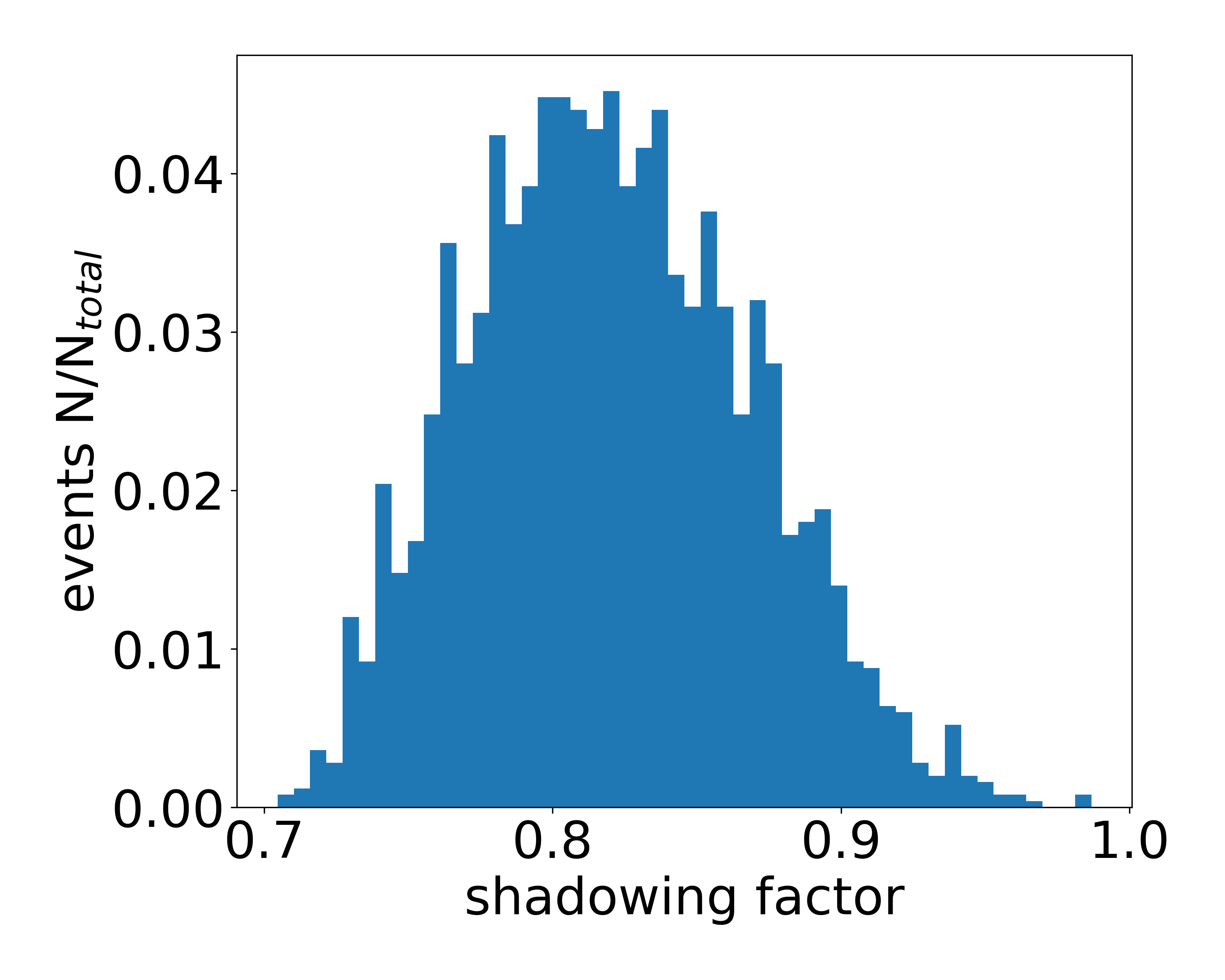}
\includegraphics[width=0.39\textwidth]{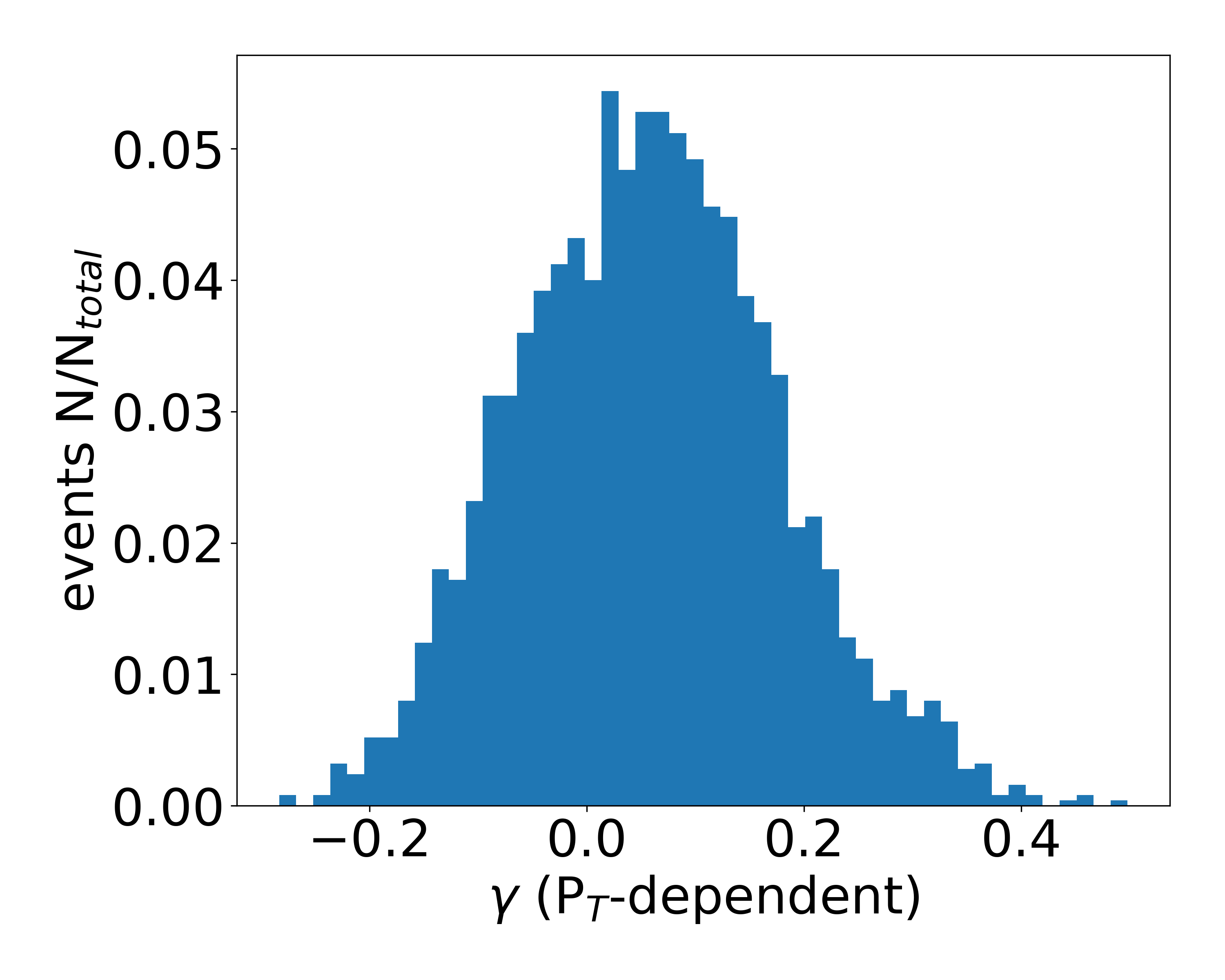}
\caption{  The distribution of parameter values $(S, \gamma)$ obtained from the CNN model is shown in the plot. The x-axis corresponds to the values of 
$S$ or $\gamma$, while the y-axis represents the number of events. 
}
\label{lab-param-S-pt}
\end{figure}

Consequently, we obtain 10K different outputs, which are plotted in Fig. \ref{lab-param-alpha}. Each data point represents an individual event, with the corresponding values of 
$\alpha$ and $\beta$ represented on the x- and y-axes, respectively. As previously mentioned, 
$\alpha$ signifies the value of 
$D_s 2\pi T$ at the critical temperature and zero momentum, while 
$\beta$ represents its temperature dependence. The majority of events are concentrated within the region 
$4\le \alpha \le 6.5$ and $0\le \beta \le 5.0$. 
Furthermore, we also plot the values of 
$\gamma$, which characterizes the transverse momentum dependence, and the shadowing factor 
$S$ arising from the cold nuclear matter effect in Fig. \ref{lab-param-S-pt}. From the distributions, it is observed that the majority of events fall within the range 
$0.0\le \gamma \le 0.2$ and $0.75\le S\le 0.9$, 
 as depicted in Fig. \ref{lab-param-S-pt}. The distribution of events in the figure reflects the uncertainty in the parameter values resulting from experimental data errors and also the neural network structure.

Based on the distributions obtained from the CNN outputs, we can directly extract the mean values of these parameters. These mean values are presented in Table \ref{tab:mean-value}. It is worth noting that the value of 
$\alpha$ remains larger than the results obtained from lattice QCD calculations at 
$T=T_c$~\cite{Altenkort:2023oms} and consistent with the values given by previous model calculations. Additionally, the temperature dependence ($\beta$) is strong, while the momentum dependence ($\gamma$) is relatively weak.
\begin{table}[htbp]
  \centering
  \caption{The mean values and the standard deviation of the parameters are extracted using the experimental data points for non-prompt  $J/\psi$ $R_{AA}(p_T)$ and $v_2(p_T)$ in 5.02 TeV Pb-Pb Collisions. }
  \label{tab:mean-value}
  \begin{tabular}{|c|c|c|}
    \hline
     parameters  & Mean values & Standard deviation\\
     \hline
     shadow factor & $\langle S\rangle = 0.82$ & 0.050\\
     \hline
     $D_s 2\pi T$ & $\langle \alpha\rangle =4.87$ & 0.90\\
     with &  & \\
     T-dependence & $\langle \beta\rangle =4.16$ & 2.32\\
     $p_T$-dependence & $\langle \gamma\rangle =0.058$ &0.12 \\
     \hline
       \end{tabular}
\end{table}

\section{summary}

In this study, the Convolutional Neural Network is employed to extract the temperature and momentum dependence in the spatial diffusion coefficient of heavy quarks using experimental data from non-prompt 
$J/\psi$ decays in B-mesons. The Langevin equation is utilized to describe the dynamical evolution of heavy quarks in the QGP and B-mesons in the hadronic gas. Additionally, the instantaneous coalescence model is used to describe the hadronization process from bottom quarks to B-mesons.
By taking different values of the shadowing factor and diffusion coefficient, the nuclear modification factors and elliptic flows of non-prompt 
$J/\psi$ in multiple centralities of 5.02 TeV Pb-Pb collisions are generated. The CNN model is trained under supervision with model calculations. To extract the values of the diffusion coefficient, we sample $(R_{AA}, v_2)$ from the experimental data points along with their error bars, which are used as inputs for the CNN. By doing so, the corresponding values of the diffusion coefficient and shadowing factor are obtained concurrently.
The dispersion in the diffusion coefficient values can be partially attributed to the uncertainties present in the experimental data. The mean values of the diffusion coefficient are also extracted. This research contributes to the understanding of the heavy quark diffusion coefficient through a data-driven analysis approach.

\vspace{1cm}
{\bf Acknowledgement: } Baoyi Chen appreciates inspiring discussions with Dr. Kai Zhou at the beginning of this work. This work is supported by the National Natural Science Foundation of China
(NSFC) under Grants No. 12175165. 


\end{spacing}
\end{document}